\documentclass[twocolumn]{aastex631}
\usepackage{amsmath}

\begin{document}

\title{Comparison of Equations of State for Neutron Stars with First-Order Phase Transitions: A Qualitative Study}

\email{anshuman18@iiserb.ac.in,\\ asim21@iiserb.ac.in,\\ mallick@iiserb.ac.in}

\author[0000-0003-1103-0742]{Anshuman Verma}
\affiliation{Indian Institue of Science Education and Research Bhopal}

\author[0009-0000-8375-4833]{Asim Kumar Saha}
\affiliation{Indian Institue of Science Education and Research Bhopal}

\author[0000-0003-2943-6388]{Ritam Mallick}
\affiliation{Indian Institue of Science Education and Research Bhopal}

\begin{abstract}
The equation of state is fundamental in describing matter under the extreme conditions characteristic of neutron stars and is central to advancing our understanding of dense matter physics. A critical challenge, however, lies in accurately modelling first-order phase transitions while ensuring thermodynamic consistency and aligning with astrophysical observations. This study explores two frameworks for constructing EoSs with first-order phase transitions: the polytropic interpolation method and the randomized speed-of-sound interpolation approach. It is found that the mass-radius relation and pressure vs. energy density relation are blind towards the thermodynamic consistency check. The polytropic interpolation method can exhibit discontinuities in the chemical potential for first-order phase transition, raising concerns regarding potential causality violations and thermodynamic inconsistencies.
In contrast, the speed of sound interpolation approach ensures continuity in the chemical potential, offering a more thermodynamically consistent and reliable framework. Moreover, the sound speed method effectively captures the softer segment of the mass-radius spectrum, a capability not achieved by the consistent piecewise-polytropic approach due to its monotonic stiffness constraints. The speed of sound definition involving number density and chemical potential reveals the thermodynamic inconsistency, making it a more consistent and robust definition. These findings underscore the importance of thermodynamic consistency in EoS construction and highlight the advantages of the randomized speed-of-sound method for modelling phase transitions in dense matter. 
\end{abstract}

\section{Introduction}
Equation of state (EoS) is fundamental for understanding matter at extreme densities, particularly within neutron stars \citep{lattimer_2004,lattimer_2014,larry_2019,bombaci_2022,Gao_2025}. The EoS is the relation between the thermodynamic variables of the system, which describe matter properties. The EoS at extreme densities is still unknown, and both ab initio calculation and earth-based experiments are still being awaited. One, therefore, relies on model calculations and finally matches their predictions with experiments \citep{schoof_2015, jorge_2017,Mandal_2020,jamie_2024}. The only laboratory to test the models is the observation from neutron stars (NSs). In the last few decades, there has been tremendous work in this regard, both theoretically and observationally \citep{zdunik_2008,Abbott_2020,prl_2022,kumar_2024}. One of the most efficient ways to constrain the EoS is constructing an ensemble of EoS by randomizing the speed of sound and using thermodynamic bounds \citep{chu_2014, Chatterjee2024,prateek_2024}. Once the ensemble is constructed, the EoS ensemble is constrained using the astrophysical observational bounds \citep{haensel_2002,lattimer_2006,miller_2021}. Initially, this was employed for smooth EoS (EoS does not have any first-order phase transition (FOPT)) and has been very successful in constraining the EoS to a large extent. Later, an ensemble of FOPT was also constructed using a piecewise-polytrope (PP) approach and then astrophysical observational bounds were employed to constrain the EoS \citep{vijaykumar_2021,mallick_2022,abbott_2022, Gorda_2023}.

EoS with FOPT signifies a rapid, discontinuous change in the internal structure of dense matter, potentially leading to quark matter cores \citep{bombaci_2000, Demorest2010}. These transitions significantly impact macroscopic properties like neutron stars' mass, radius, and frequencies \citep{kl_2007,wang_2012,yang_2017}. The EoS also influences the NS stability and observable properties like star-quakes and magnetic field dynamics \citep{Bhatt_2017}. A PT may also result in abrupt changes in radius or mass, which can produce detectable signals in multi-messenger astrophysics \citep{alford_2007,zheng_2016,miller_2021}.
Identifying signatures of FOPT is critical for advancing our understanding of compact objects and quantum chromodynamics (QCD) \citep{bissell_2014,houjun_2015,Abbott_2020}. However, astrophysical observation to date has not been able to clearly distinguish whether quark matter exists at the cores of neutron stars, let alone whether there is a smooth or FOPT \citep{Xujun_2009,acernese2014,aasi_2015,bauswein2019}. 

Model construction of an ensemble of EoS is a novel idea; however, one should be careful and check its thermodynamic consistency. As these are not microscopic models, one should check them rigorously so that they do not violate thermodynamical consistency. More specifically, FOPT does have a discontinuity in energy and number density; however, pressure and chemical potential must be continuous. There lies a pitfall as the construction of NSs mainly involves energy density, number density and pressure. The chemical potential is not directly related to the Tolmann-Volkoff-Oppendeimer (TOV) \citep{TOV}. The PP approach \citep{Gorda_2023} does not effectively spell out the chemical potential continuity and the mass-radius curves are blind to it. For a correctly modelled FOPT EoS, the pressure and chemical potential must be continuous at the transition point. This ensures mechanical and thermodynamic equilibrium, ensuring Euler theorem and the Gibbs phase rule \citep{Radhakrishnan1990-pc,jana2018}. The first-order transition is characterized by a discontinuous jump in density (or volume per particle) as the system shifts from a lower-density phase (e.g., hadronic matter) to a higher-density phase (e.g., quark matter) or vice-versa \citep{bauswein_2019,zhou_2024}.

The speed of sound interpolation (SoSI) is not usually used to generate FOPT and is restricted only to generate smooth EoS; it can also be employed to generate FOPT. The main aim of this paper is to generate thermodynamic consistent EoS using the two formalisms: the SoSI and PP. More importantly, the condition that needs to be fulfilled to construct a correct thermodynamic consistent EoS using the PP approach is derived. The structure of the paper is organized as follows: Section \ref{sec:2} addresses the construction of the EoS ensemble, ensuring thermodynamic consistency. Section \ref{sec:4} presents the results, and Section \ref{sec:5} summarises our findings and concludes the study.

\section{Construction Methods}\label{sec:3}

The construction methods addressed in this section are:
\begin{itemize}
    \item Piecewise-polytropic (PP) method (\cite{Gorda_2023}).
    \item Piecewise-polytropic consistent (PP-consistent) method.
    \item Speed of sound interpolation (SoSI) method (\cite{Annala2020}).
\end{itemize}
The PP method suffers a chemical potential jump across the transition, which leads to inconsistencies, as discussed in section \ref{sec:4}. We propose the PP-consistent method to address the inconsistency, which ensures a continuous chemical potential across the jump.

\subsection{Piecewise Polytropic (PP) Construction}\label{sec:pp}

An ensemble of EoS using a PP approach is constructed up to a baryon density of 10$n_0$ (where $n_0 = 0.16 \, \text{fm}^{-3}$, nuclear saturation density).%\textcolor{black}{with the high-density constraints enforced through matching with pQCD calculations}.
The construction of the EoS involves three key steps:

\begin{enumerate}
    \item For densities $n \leq 0.57n_0$, we employ the BPS crust EoS from \citet{Baym_eos}. From $n = 0.57n_0$ up to $n_{\text{CET}} = 1.1n_0$, we interpolate between the soft and stiff EoSs within the chiral CET band of \citet{Hebeler_2013}.
    
    \item For densities $n_{\text{CET}} < n \leq n_{\text{PT}}$, where $n_{PT} \in [1.1,10] n_0$ ; a single polytropic form is used: 
    \[
    p(n) = p_{\text{CET}}(n_{\text{CET}}) \left(\frac{n}{n_{\text{CET}}}\right)^{\Gamma_1}
    \]
    
    \item For $n_{\text{PT}} + \Delta n < n \leq 10n_0$; where $\Delta n/n_{PT} \in [0,9.09]$, a second polytropic EoS is applied: 
    \[
    p(n) = p(n_{\text{PT}})\left(\frac{n}{n_{\text{PT}} + \Delta n}\right)^{\Gamma_2}
    \]
\end{enumerate}
The phase transition (PT) occurs between the two polytropic segments. It is characterized by two key parameters: the onset of the phase transition \( n_{PT}\) and the density jump \( \Delta n\).

\subsection{Speed-of-Sound Interpolation (SoSI) Construction }
\begin{figure}[h]
    \centering
    \includegraphics[width=0.99\linewidth]{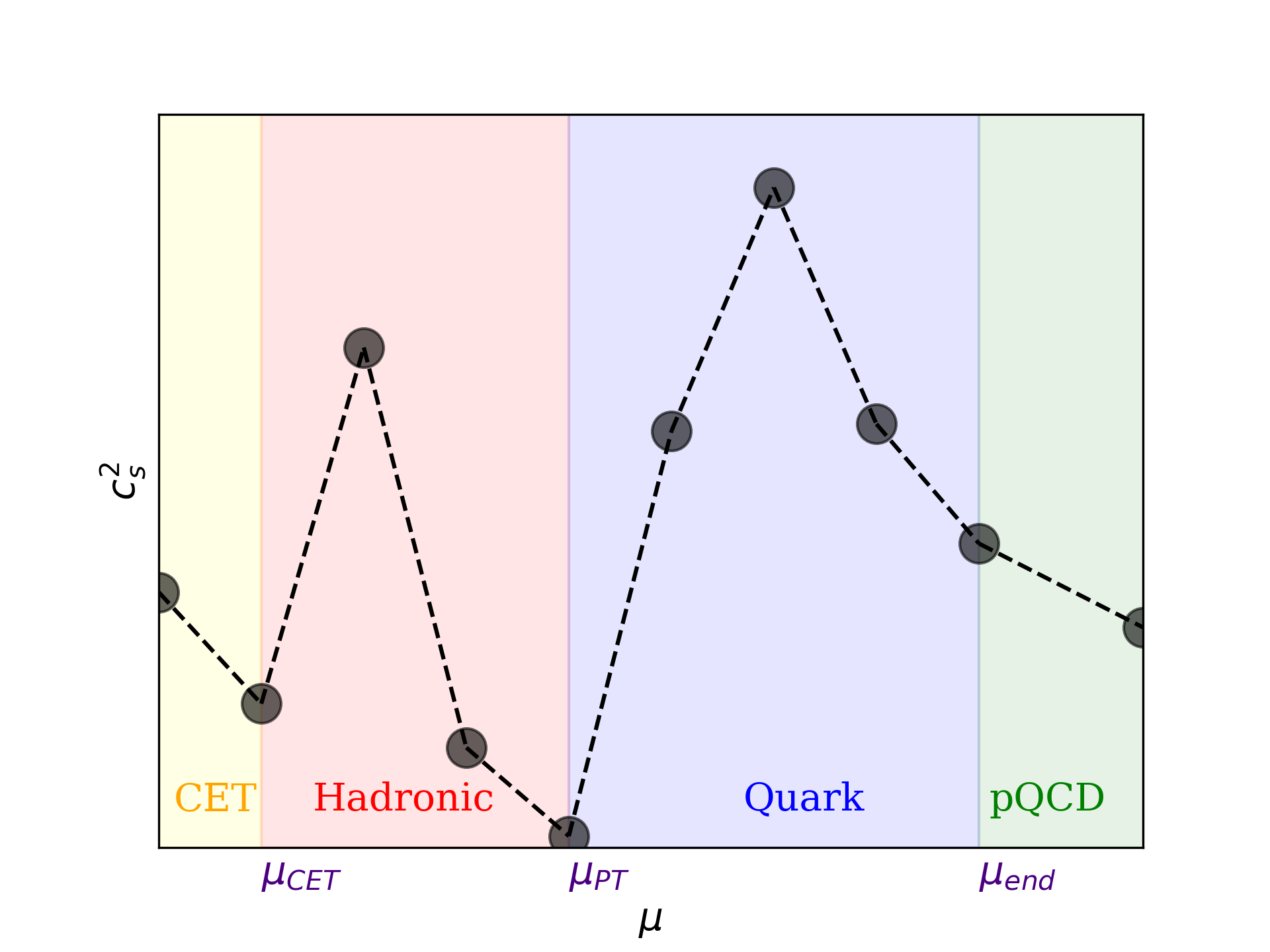}
    \caption{Diagram illustrating the speed of sound randomization in the $c_s^2 - \mu$ space. A first-order phase transition happens at $\mu_{PT}$, and the pQCD region begins at $\mu_{end} = 2.6$ GeV.}
    \label{fig:schematic}
\end{figure}
The speed of sound ($c_s^2$) is parametrised as a function of chemical potential ($\mu$) \citep{Annala2020}, which allows us to write the number density as 
\begin{equation}
n(\mu) = n_{\scriptscriptstyle CET}\exp\left[\int_{\mu_{CET}}^{\mu} \frac{d\mu'}{\mu' c_s^2(\mu')}\right]
\label{1}
\end{equation}
where $n_{CET}$ is fixed by matching the lower-density CET EoS to the initials of the interpolated EoS, and $\mu_{CET}$ is the chemical potential corresponding to $n_{CET}$. Pressure can then readily be calculated as
\begin{equation}
p(\mu) = p_{\scriptscriptstyle CET} \! + n_{\scriptscriptstyle CET}\int_{\mu_{CET}}^{\mu}d\mu' \exp\left[\int_{\mu_{CET}}^{\mu'}\frac{d\mu''}{\mu''c_s^2(\mu'')}\right]
\label{2}
\end{equation}   
where $p_{CET}$ corresponds to the pressure value at the matching point of the CET EoS. Our study here includes 5 randomised segments of ($c_{s,i}^2,\mu_{i}$), where $\mu_{i} \in [\mu_{\scriptscriptstyle CET},2.6$ $GeV]$. The upper limit of $\mu$ is chosen by Fraga et al. \citep{Fraga_2014}, such that the uncertainty in the pQCD regime is roughly the same as the uncertainty at CET. A piecewise linear function is employed to connect the points $\left\{\mu_{i},c_{s,i}^{2}\right\}$ as:
\begin{equation}
c_s^2(\mu) = \frac{(\mu_{i+1} - \mu)c_{s,i}^2 + (\mu - \mu_{i})c_{s,i+1}^2}{\mu_{i+1} - \mu_i}
\label{3}
\end{equation}
where $c_{s,i}^2 \in [0,1]$. This function aids us to carry out the integrals of \eqref{1} and \eqref{2}.

SoSI construction method inherently permits FOPT; however, they are statistically suppressed. To address this, we modify the construction by dividing it into two segments: the first and second branches (see Fig \ref{fig:schematic}). The first branch, which mimics the hadronic branch, is extended up to a chemical potential of $\mu_{\scriptscriptstyle PT}$; where $\mu_{\scriptscriptstyle PT}$ is chosen randomly in the range between $\mu_{\scriptscriptstyle CET}$ and onset of pQCD potential. In order to induce a FOPT between the two branches, the first branch's endpoint and the second branch's starting point are particularly significant. The integrals for the thermodynamic quantities at the endpoint of the first branch are expressed as:
\begin{align}
{n(\mu_{\scriptscriptstyle{PT}})} &= {n_{\scriptscriptstyle CET}\exp\left[\int_{\mu_{\scriptscriptstyle CET}}^{\mu_{\scriptscriptstyle PT}} \frac{d\mu'}{\mu' c_s^2(\mu')}\right]} \label{eq:4}\\
{p(\mu_{\scriptscriptstyle{PT}})} &= {p_{\scriptscriptstyle CET} \! + n_{\scriptscriptstyle CET}\int_{\mu_{\scriptscriptstyle CET}}^{\mu_{\scriptscriptstyle PT}}d\mu' \exp\left[\int_{\mu_{\scriptscriptstyle CET}}^{\mu'_{\scriptscriptstyle PT}}\frac{d\mu''}{\mu''c_s^2(\mu'')}\right]}
\end{align}

To exhibit FOPT, the number density at the starting point of the second branch is induced with a jump of $\Delta n$, which allows us to write the integrals of \eqref{1} and \eqref{2} for the thermodynamic quantities of the second branch as:
\begin{align}
{
    n(\mu)} &= {(n_{\scriptscriptstyle PT} + \Delta n )\exp\left[\int_{\mu_{PT}}^{\mu} \frac{d\mu'}{\mu' c_s^2(\mu')}\right]} \\
    {p(\mu)} &= {p_{\scriptscriptstyle PT} \! + (n_{\scriptscriptstyle PT} + \Delta n)\int_{\mu_{PT}}^{\mu}d\mu' \exp\left[\int_{\mu_{PT}}^{\mu'}\frac{d\mu''}{\mu''c_s^2(\mu'')}\right]}
\end{align}
where $p_{\scriptscriptstyle PT} = p(\mu_{\scriptscriptstyle PT})$ and $n_{\scriptscriptstyle PT} = n(\mu_{\scriptscriptstyle PT})$. The value of $\Delta n$ is chosen to span the space similar to that of the polytropic construction. The parameters defining FOPT for the SoSI construction are $\mu_{PT}$ and $\Delta n$, while for the PP construction, they are $n_{PT}$ and $\Delta n$. The stiffness of the two branches in the PP construction is determined by the parameters $\Gamma_1$ and $\Gamma_2$, while the SoSI construction is governed by the speed of sound segments, which are randomized.
Adhering to the astrophysical constraints of \cite{Demorest2010}, we ensure that every EoS is capable of generating M$_{\scriptscriptstyle TOV}>2.0 M_\odot$ as well as respecting the tidal deformability limit of \(\Tilde{\Lambda} < 720\). 

This study analyses the EoS derived from polytrope-based construction and the SoSI method. In order to facilitate a valid comparison between the two methodologies, we ensure that the transition onset density for the SoSI method is maintained within the interval $[1.1,10] n_0$, which aligns with the range employed in the polytropic construction method. This alignment is achieved by imposing a constraint on the value of the integral in equation \ref{eq:4}, keeping it below the threshold of $10n_0$. 

\subsection{Piecewise Polytrope Consistent (PP-consistent) method}\label{sec:2}
To confirm the continuity of the chemical potential within the framework of the polytropic construction method, as elaborated in \ref{sec:pp}, we define the endpoint thermodynamic variables of the hadronic branch as 
\begin{align}
    p^{\scriptscriptstyle{h}}_{\scriptscriptstyle{PT}}(n^{\scriptscriptstyle{h}}_{\scriptscriptstyle{PT}}) &= p_{\scriptscriptstyle{CET}}\left(\frac{n^{\scriptscriptstyle{h}}_{\scriptscriptstyle{PT}}}{n_{\scriptscriptstyle{CET}}}\right)^{\Gamma_1}, \notag \\
    \epsilon^{\scriptscriptstyle{h}}_{\scriptscriptstyle{PT}}(n^{\scriptscriptstyle{h}}_{\scriptscriptstyle{PT}}) &= m_{\scriptscriptstyle{b}} n_{\scriptscriptstyle{PT}}^{\scriptscriptstyle{h}} + \frac{p^{\scriptscriptstyle{h}}_{\scriptscriptstyle{PT}}}{\Gamma_1 - 1}, \notag \\
    \mu^{\scriptscriptstyle{h}}_{\scriptscriptstyle{PT}} &= m_{\scriptscriptstyle{b}} + \frac{p^{\scriptscriptstyle{h}}_{\scriptscriptstyle{PT}}}{n^{\scriptscriptstyle{h}}_{\scriptscriptstyle{PT}}}\left(\frac{\Gamma_1}{\Gamma_1 - 1}\right) 
    \label{eq:had_pt}
\end{align}

After undergoing FOPT, at the onset of the quark branch, the thermodynamic parameters are expressed as:
\begin{align}
    \epsilon^{\scriptscriptstyle{q}}_{\scriptscriptstyle{PT}}(n^{\scriptscriptstyle{q}}_{\scriptscriptstyle{PT}}) &= m_{\scriptscriptstyle{q}} n_{\scriptscriptstyle{PT}}^{\scriptscriptstyle{q}} + \frac{p^{\scriptscriptstyle{q}}_{\scriptscriptstyle{PT}}}{\Gamma_2 - 1}, \notag \\
    \mu^{\scriptscriptstyle{q}}_{\scriptscriptstyle{PT}} &= m_{\scriptscriptstyle{q}} + \frac{p^{\scriptscriptstyle{q}}_{\scriptscriptstyle{PT}}}{n^{\scriptscriptstyle{q}}_{\scriptscriptstyle{PT}}}\left(\frac{\Gamma_2}{\Gamma_2 - 1}\right) 
    \label{eq:quark_pt}
\end{align}
where the pressure at this junction is maintained as $p_{\scriptscriptstyle{PT}}^{\scriptscriptstyle{q}} = p_{\scriptscriptstyle{PT}}^{\scriptscriptstyle{h}}$. The number density adapts according to $n_{\scriptscriptstyle{PT}}^{\scriptscriptstyle{q}} = n_{\scriptscriptstyle{PT}}^{\scriptscriptstyle{h}} + \Delta n$. By equating the chemical potentials given by equations \ref{eq:had_pt} and \ref{eq:quark_pt}, the thermodynamic consistency condition reduces to
\begin{equation}
    \frac{(m_q - m_b)n_{\scriptscriptstyle{PT}}^{\scriptscriptstyle{q}}}{p_{\scriptscriptstyle{PT}}}\frac{(\Gamma_1 - 1)}{\Gamma_1} + \frac{\Gamma_1 - \Gamma_2}{\Gamma_1(\Gamma_2 - 1)} = \frac{\Delta n}{n_{\scriptscriptstyle{PT}}^{\scriptscriptstyle{h}}}
    \label{eq:10}
\end{equation}
This relationship is essential to ensure that the chemical potential remains continuous during the transitional phase. By imposing this condition, one can operate within the framework of the piecewise polytropic formalism, guaranteeing thermodynamic consistency across the transition.

%%%%%%%%%%%%%%%%%%%%%%%%%%%%%%%%%%%%%%%%%%%%%%%%%%%%%%%%%%%%%%%%%%

\section{Results and Discussion}\label{sec:4}

\begin{figure*}
    \centering
    \includegraphics[width=0.99\linewidth]{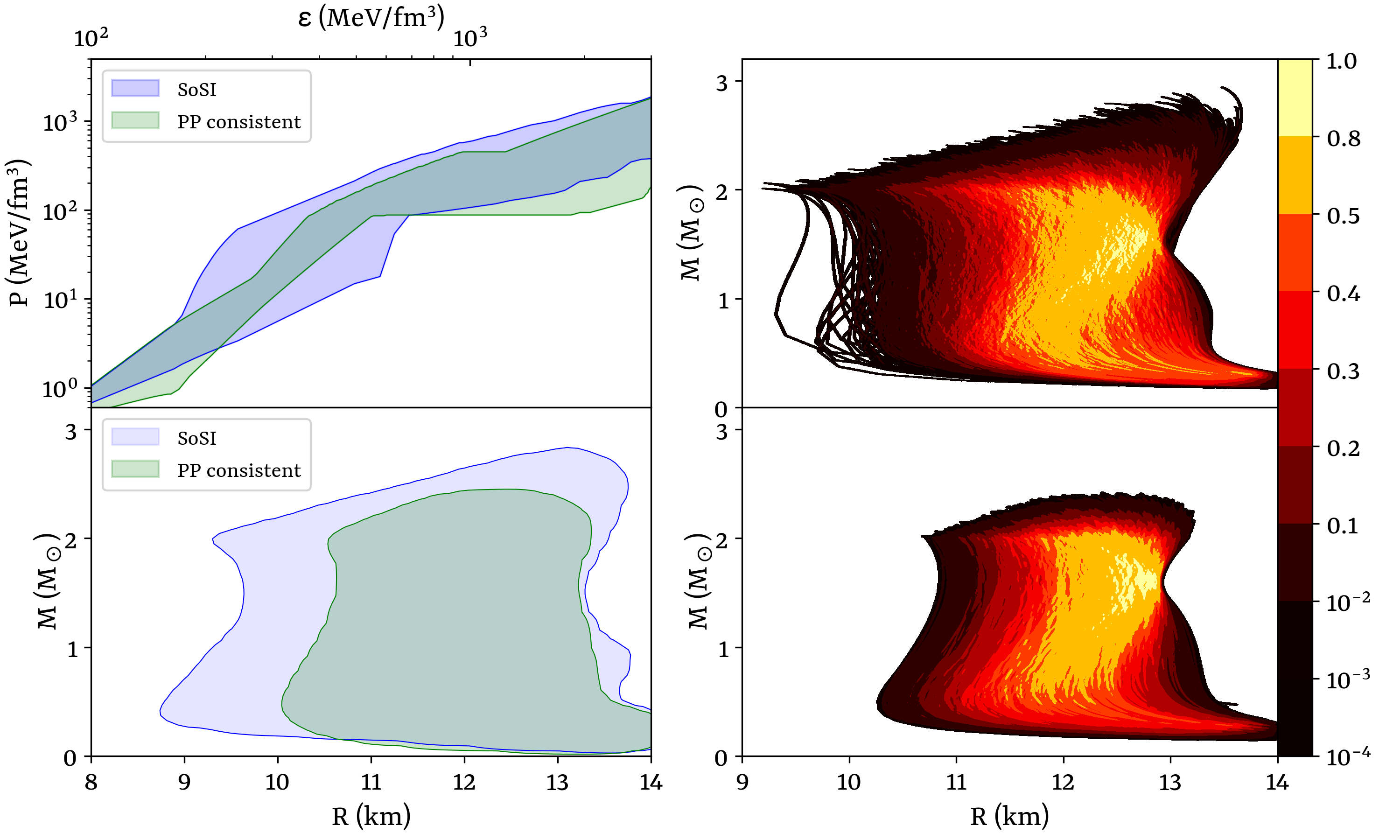}
    \caption{The right panel compares the EoS (top) and the corresponding M-R contour (bottom), achieved from the two distinct construction methods: SoSI method as well as PP-consistent method. The grey contour refers to the permissible M-R contour from \cite{Annala2020}, which does not incorporate explicit phase transitions. The left panel shows the population density of stars, achieved from the ensemble of EoS with SoSI construction (top) and PP-consistent construction (bottom).}
    \label{combined}
\end{figure*}

The EoS constructed in two different ways are considerably different.
Figure \ref{combined} presents a comparative analysis of the EoS and mass-radius diagram obtained from two different methods.
In the left (top) panel, the plot illustrates the relationship between pressure (\(P\)) and central energy density (\(\epsilon\)) on a logarithmic scale for EoS models created using the PP-consistent method (green contour) and the SoSI method (blue contour). Both approaches comply with causality constraints. The PP-consistent method employs predefined functional forms, resulting in a more constrained FOPT EoS. This leads to a narrower range of possible behaviours, with controlled variations at intermediate densities and convergence toward the pQCD limit at high densities.
Conversely, the SoSI method (blue contour) offers greater flexibility, allowing a broader spectrum of phase transition scenarios. The wider spread of EoS curves at lower energy densities (up to approximately 1000 MeV/fm³) highlights the uncertainties associated with nuclear matter properties. However, this variability diminishes at higher densities as the EoS approaches the pQCD limit. This broader distribution at low densities reflects the sensitivity of the speed of sound to phase transition characteristics, capturing a wider range of potential EoS behaviours. The left (bottom) panel displays the corresponding mass-radius (M-R) curves obtained by solving the TOV equations. Both methods satisfy key astrophysical constraints, including a maximum neutron star mass exceeding \(2M_\odot\) and a tidal deformability limit of \(\widetilde{\Lambda}< 720\). The PP-consistent method (green contour) produces more tightly clustered M-R curves, indicating a more constrained range of neutron star properties. In contrast, the SoSI method (blue contour) exhibits a broader spread, reflecting more significant variability in phase-transition behaviour \citep{Annala2020}.

The right panel of Fig. \ref{combined} presents the PDF plot of the M-R curves for the SoSI construction method (top) and the PP-consistent method (bottom). While the SoSI construction covers a larger area in the M-R space, it is mainly concentrated in a more confined region. A similar pattern is observed for the PP-construction method, with the concentration increases in a confined region having a substantial overlap in coverage with the SoSI ensemble.

For early phase transitions, generally happening at a density less than 2$n_0$, a substantial increase in stiffness beyond the transition is necessary to meet the constraints set by observations (simultaneous satisfaction of the $2M_\odot$ bound and also maintaining the limit of tidal deformability $\Lambda < 720$). This requirement imposes stringent limits on the EoS and the corresponding M-R diagram. As both the PP construction methods are restricted compared to the SoSI method, they could not generate consistent EoS which satisfies astrophysical bounds. It is most interesting to note that although PP (inconsistent) was able to generate at least a few EoS satisfying the astrophysical bound, the PP-consistent method was not able to generate any (detailed in the Appendix section \ref{incon}). 
For the PP-consistent approach, all the EoS undergo transitions at densities greater than $2n_0$. The limitation of the PP-consistent method arose because of the consistency condition (eqn \ref{eq:10}). For the PP method, $\Gamma_2$ is considered a free parameter, offering more flexibility in stiffness adjustment. However, within the PP-consistent methodology, $\Gamma_2$ is not independently variable, as its value is directly influenced by the parameters $\Gamma_1$, $\Delta n$, and $n^h_{\scriptscriptstyle{PT}}$ as specified by equation \ref{eq:10}. On the other hand, the SoSI method does not have such restriction and allows for significant variation in stiffness even after the early PT to satisfy astrophysical bounds. 

A particularly noteworthy observation is that the PP-based EoS cannot cover the softer region of the M-R spectrum. This arises from the functional nature of polytropes, which exhibit monotonically increasing stiffness. Consequently, a single polytrope cannot simultaneously produce the required soft segment to describe low-mass stars; thereafter, the stiff segment is needed to satisfy astrophysical constraints \citep{Ferrari, Kurkela_2014}.

\begin{figure*}
    \centering
    \includegraphics[width=0.49\linewidth]{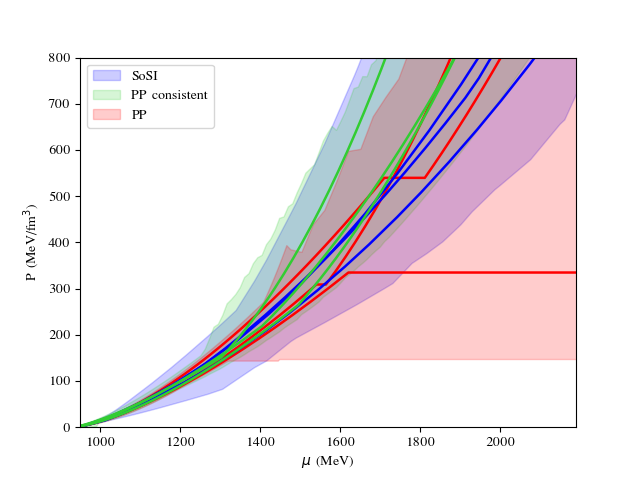}
    \includegraphics[width=0.49\linewidth]{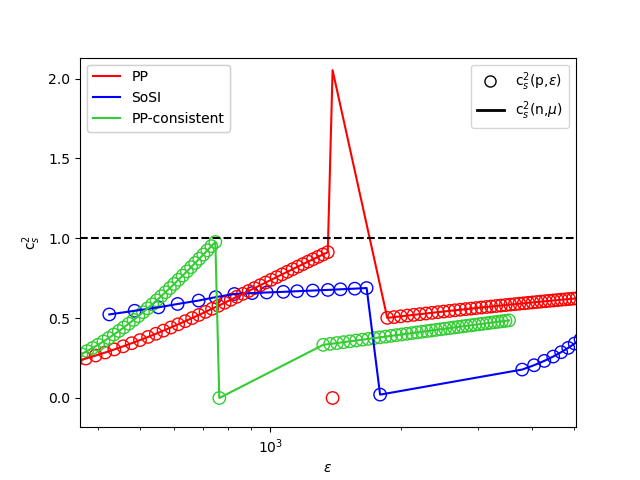}
    \caption{The plot on the left displays the relationship between chemical potential (\(\mu\)) and pressure (\(P\)) for three different EoS models. The EoS constructed using the PP method (depicted in red) shows a discontinuity in the \(\mu\)-\(P\) curve due to specific aspects of its formulation. In contrast, the EoS derived from the SoSI and the PP-consistent approach produces a smooth, continuous curve (depicted in blue and green respectively). The discontinuity in the PP EoS results in a causality violation, as the computed speed of sound exceeds the causal limit when determined from the chemical potential and number density, which serves as our check to determine consistent EoS. Such a violation implies that the PP EoS is physically inconsistent. Conversely, the SoSI-based EoS, as well as PP-consistent maintains causality, as shown in the right panel, where the speed of sound remains within the causal threshold, indicating a more consistent and valid model.}
    \label{chem-figure}
\end{figure*}

As already stated, it is difficult to infer the consistency of the FOPT EoS from $P$ vs. $\epsilon$ curve or the M-R sequence curve. One of the most essential thermodynamic criteria, even for FOPT, is that the pressure and chemical potential should be smooth without any discontinuity. 
The left panel of Figure \ref{chem-figure} illustrates the relationship between \(\mu\) and \(P\) for three distinct EoS models. The EoS constructed using the PP method (depicted in red) exhibits a discontinuity in the \(\mu\)-\(P\) curve, which stems from the specific formulation of this approach. In contrast, the SoSI-based EoS (blue) and PP-consistent method (green) produce a smooth, continuous curve. 

A critical parameter for identifying phase transitions in dense matter is \(c_s^2\). This can be defined in two ways: in terms of pressure and energy density (\(c_s^2(P, \epsilon)\)) or using number density and chemical potential (\(c_s^2(n,\mu)\)) \citep{Abgaryan}. 

\begin{equation}
    c_s^2 = \left( \frac{\partial (log_{e}\mu)}{\partial (log_{e} n)} \right)_s = \left( \frac{\partial P}{\partial \epsilon} \right)_s \label{eq:sos}
\end{equation} 

Both definitions should maintain the causality for an EoS to be physically consistent. In the case of the PP-based model, the discontinuity in the \(\mu\)-\(P\) relationship manifests itself as a discrepancy between these two relations. As confirmed by fig \ref{chem-figure} right, the SoSI method EoS maintains causality and generates the same curve with two different definitions. Whereas the PP-consistent EoS maintains causality and generates the same curve for two different definitions, the same is not true for inconsistent PP EoS. The speed-of-sound curve differs, and the definition involving chemical potential and number density violates causality. The figure also hints at one of the pitfalls of the consistency check of the EoS, as the speed of sound definition with pressure and energy density does not violate causality and seems to be thermodynamically consistent.  

The discrepancy in the M-R diagram (not having stars with consistent PP EoS for transition density less than twice the saturation density) results from this discrepancy. The PP method is a more restricted method, only using a single polytrope before and after the PT. If more than one polytrope is used, one can have more freedom in obtaining more EoS, which satisfies both thermodynamic consistency and astrophysical bound. This reinforces the conclusion that the SoSI method offers a more comprehensive and reliable framework for modelling first-order phase transitions in dense neutron star matter.

\section{Summary and Conclusion}\label{sec:5}

The study examines the construction of an ensemble of EoS involving FOPT. Two separate methods are used for the construction: the SoSI and the PP method. One of the paper's key findings is that the $P$ vs. $\epsilon$ diagram or the $M-R$ curve are blind towards the consistency check of the EoS. A consistent EoS, even with FOPT, should not have any pressure or chemical potential discontinuity. Even if they follow the low and high-density limits and satisfy the astrophysical bounds of neutron stars and GW170817 tidal deformability, the EoS can be thermodynamically inconsistent. 

The study identifies the consistency condition for the PP ensemble of EoS. However, even with consistent EoS, the PP-consistent EoS and SoSI method produces very different EoS ensembles. The SoSI method has more flexibility and can generate all possible EoS models, whereas even PP-consistent EoS is largely restricted. The PP-consistent method has a single polytrope on either side of the PT discontinuity and, therefore, becomes very restricted in adhering to the consistency of the astrophysical constraints.

One of the essential aspects of the study is that early phase transitions at densities below \(2n_0\) are associated with larger neutron star radii but necessitate significant EoS stiffening to support \(2M_\odot\) stars. Many early transition models fail to satisfy tidal deformability constraints, particularly the limit \(\widetilde{\Lambda}<720\). This limitation is highlighted in the PP method, which restricts their ability to achieve the required balance of softness and stiffness to meet astrophysical criteria. 

The inconsistency of the PP model is also revealed in the definition of speed of sound. The speed of sound can be expressed as a function of pressure and energy density or as a function of number density and chemical potential. For inconsistent PP EoS, the definition of $c_s^2$ with energy density and pressure maintains causality, whereas it violates causality for the definition involving number density and the chemical potential. For consistent EoS, both definitions give the same $c_s^2$ maintaining causality. Therefore, one of the key points of the study is that the definition of $c_s^2$ involving $n$ and $\mu$ is more robust and should be rigorously used.

The main research contribution is the qualitative comparison of two widely used methods for constructing EoS with phase transitions. By demonstrating the critical role of chemical potential continuity and thermodynamic consistency, this work identifies the SoSI method as a more physically reliable approach for modelling dense matter in neutron stars. Future directions include expanding the analysis to include multi-messenger data to improve the robustness of these models further and provide deeper insights into the nature of dense matter and phase transitions.  

\section*{Acknowledgements}

The authors thank the Indian Institute of Science Education and Research Bhopal for providing all the research and infrastructure facilities. AV acknowledges the Prime Minister's Research Fellowship (PMRF), Ministry of Education, Government of India, for a graduate fellowship, and RM acknowledges the Science and Engineering Research Board (SERB), Govt. of India, for financial support in the form of a Core Research Grant (CRG/2022/000663).

\bibliographystyle{apsrev4-2-author-truncate}
\bibliography{references}

\appendix
\section{Thermodynamic Consistency}
The chemical potential \( \mu \), is related to the Gibbs free energy per particle \( G \) as,
\[
G = F + PV
\]
where $F$ is Helmholtz free energy, $P$ is the pressure, V is the volume. For a system with particle number \( N \), it can also be written as,
\[
G = \mu N
\]
which, by the Euler theorem, shows that the Gibbs free energy per particle is equivalent to the chemical potential \( \mu \) \cite{Mindel1962-wi}. The partial derivative of the chemical potential with respect to pressure at constant temperature is,

\begin{equation}
    \left( \frac{\partial \mu}{\partial P} \right)_T = \frac{V}{N} = v \label{eq:1}
\end{equation}
where \( v \) is the volume per particle. Since \( v \) is positive by definition, \( \mu \) must be a strictly increasing function of pressure at a constant temperature. So, for a stable thermodynamic system, the Gibbs free energy is a strictly concave function of pressure. This concavity ensures that there is thermodynamic stability in equilibrium. In the context of a first-order phase transition, this concavity implies that the chemical potentials of two homogeneous phases near the transition point intersect. At the crossing point, the phase with the lower Gibbs free energy (or chemical potential) is the more stable, representing the minimum Gibbs potential for the system \citep{Radhakrishnan1990-pc, Blundell,jana2018}.

\section{M-R curve for transition below 2$n_0$}\label{incon}

\begin{figure}[h]
    \centering
    \includegraphics[width=0.5\linewidth]{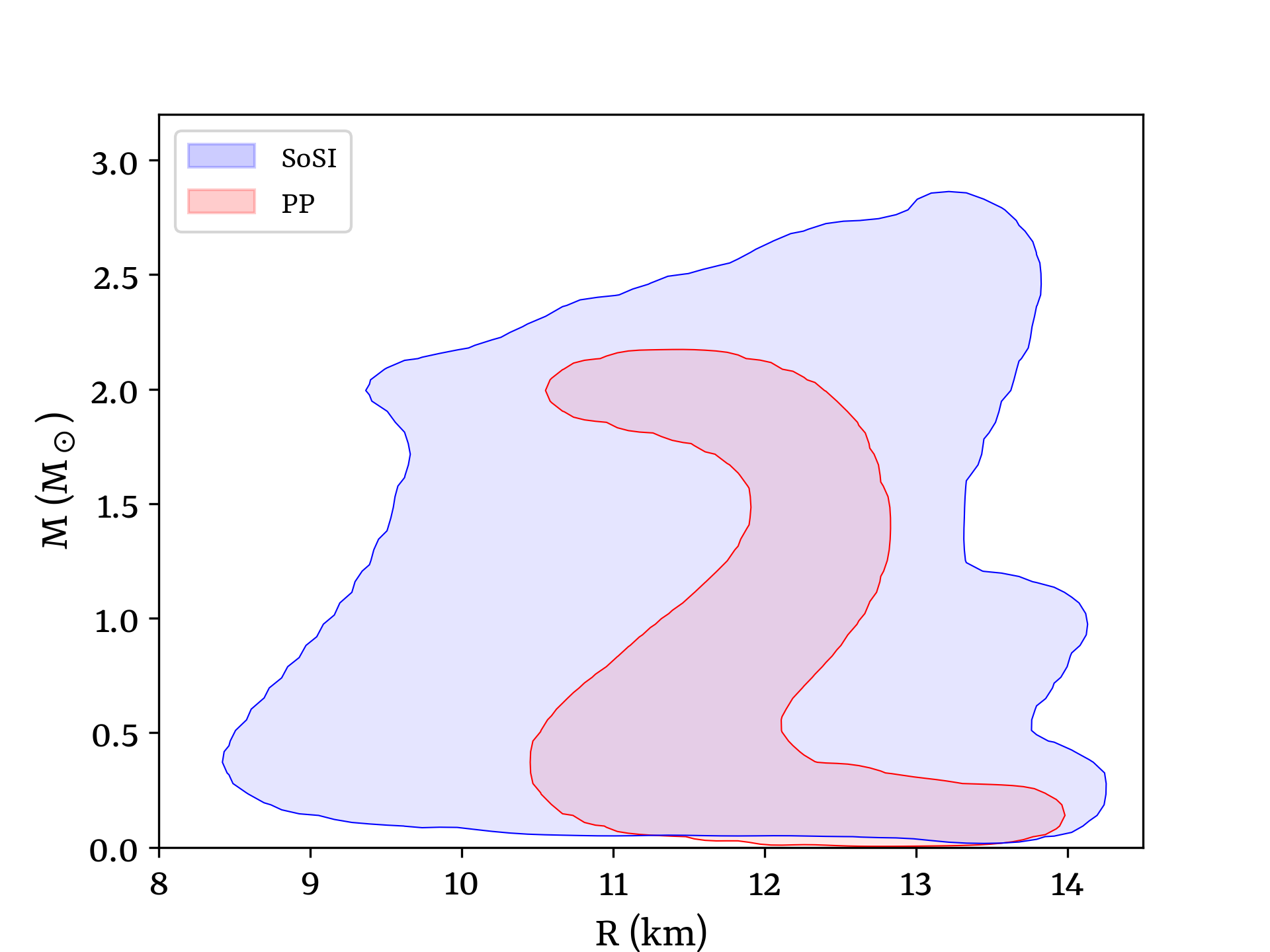}
    \caption{The MR contour derived from the SoSI method (blue contour) and PP method (red contour), showing transition onset at densities below 2$n_0$.}
    \label{trans_2n0}
\end{figure}

Figure \ref{trans_2n0} demonstrates the region within the M-R space that is spanned by the SoSI method, compared to the region spanned by the PP (inconsistent) method, for PT occurring at densities below 2$n_0$. The M-R contour generated from the sequences satisfies the astrophysical bounds of M$>2$M$_\odot$ and the tidal deformability bound of $\Tilde{\Lambda}<720$. Notably, the SoSI method encompasses an area similar to that associated with transitions above $2n_0$. In contrast, the inconsistent PP method exhibits a more confined contour due to its inherently polytropic characteristics. This results in an even more limited region within the space. Also, the PP-consistent method fails to generate any EoS when PT happens below $2n_0$. 

%\newpage
\section{Comparison with PP method}

In this section, we compare the EoS ensemble obtained using SoSI, PP-consistent and PP (inconsistent) methods. We also show the M-R region, encompassed by three different construction methods.
In the left panel, the plot illustrates the relationship between pressure (\(P\)) and central energy density (\(\epsilon\)) on a logarithmic scale. The SoSI method (blue contour) offers greater flexibility, allowing a broader spectrum of phase transition scenarios. The wider spread of EoS curves at lower energy densities (up to approximately 1000 MeV/fm³) highlights the uncertainties associated with nuclear matter properties. This broader distribution at low densities reflects SoSI capturing a wider range of potential EoS behaviours. The polytropic method (PP and PP-consistent) employs predefined functional forms, producing more tightly clustered M-R curves and indicating a more constrained range of neutron star properties.

\begin{figure*}[h]
    \includegraphics[width = 0.49\linewidth]{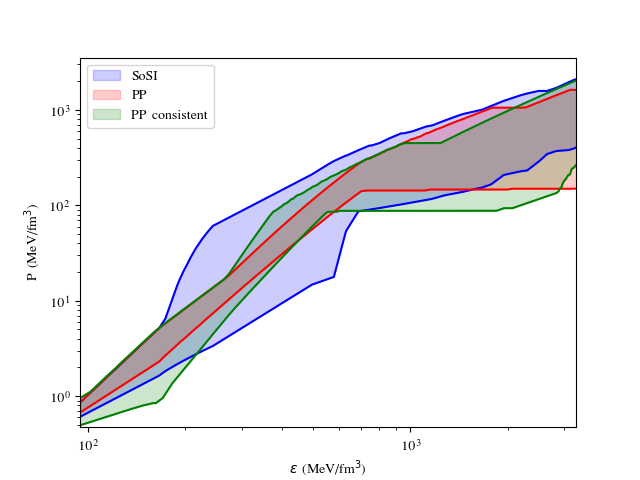}
    \includegraphics[width = 0.49\linewidth]{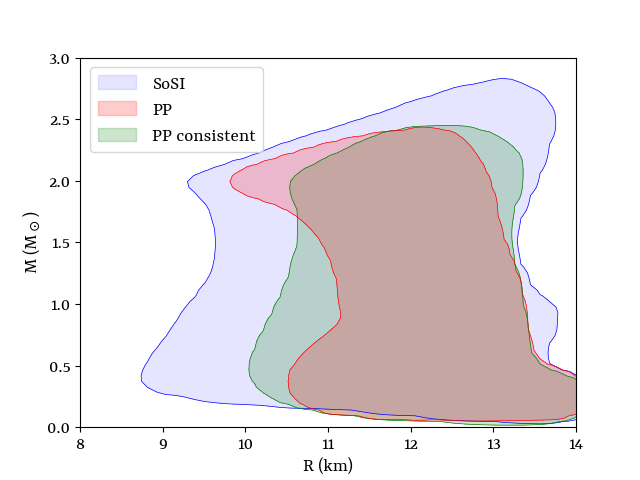}
    \caption{Left: The contour of neutron star EoS in the P-$\epsilon$ space, constructed using the SoSI method (blue), PP method (red) and PP-consistent (green) on a logarithmic scale. Right: The M-R contours achieved from the three different construction methods satisfy the astrophysical constraints of $M_{TOV} > 2 M_{\odot}$ and tidal deformability of $\Tilde{\Lambda} < 720$.}
    \label{3comparison}
\end{figure*}

\end{document}